# Structural Coloration of Transmission Light through Self-Aligned and Complementary Plasmonic Nanostructures

Myeong-Su Ahn,[†] Taerin Chung,[†] and Ki-Hun Jeong*[†]

[†]Department of Bio and Brain Engineering, Korea Advanced Institute of Science and Technology (KAIST), 291 Daehak-ro, Yuseong-gu, Daejeon 34141, Republic of Korea

* Phone +82.42.350.4323 (Office), FAX +82.42.350.4310, E-mail: kjeong@kaist.ac.kr

ABSTRACT

Structural coloration of natural surfaces often originates from the change of reflected colors depending on the viewing or illumination angle. Recently, the structural coloration of nanoplasmonic structures have attracted a great deal of attention due to high compactness, robust stability and high color-tunability, as well as high sensitivity to an incident angle. Here we report complementary plasmonic structures (CPS) for transmission structural coloration by tailoring a single spectral peak depending on the incident angle of light. The CPS features self-aligned silver nanohole and nanodisk arrays, supported by dielectric nanopillar arrays of hydrogen silsesquioxane. Unlike conventional hybridized nanostructures of plasmonic nanohole and nanodisk arrays, the nanodisks of CPS effectively attenuate undesired spectral peaks of nanoholes by exploiting an extinction peak of nanodisks, serving as a spectral suppressor. As a result, a single transmission spectral peak becomes red-shifted from 736 nm to 843 nm as the incident angle varies from 0° to 30°. This unique configuration provides a new direction for tunable filters that can be utilized for compact multispectral or hyperspectral imaging applications.

TEXT

Nature provides a myriad of color reservoirs. A structural color among the intriguing natural colors is produced by reflected light from a material surface.[1] For instance, the wing scales of C. *gloriosa* beetles and *Morpho* butterflies exhibit structural coloration depending on the change of illumination or viewing angle.[2, 3] Many previous works on the structural coloration have been experimentally demonstrated by employing nanostructures such as Bragg stack photonic crystals,[4] close-packed spheres,[5] inverse opal structures,[6] or regularly arranged concavities.[7] This structural coloration often results from constructive interference of reflected light from highly ordered array structures or multi-stacked layers of alternating high and low refractive indices. However, the reflection type photonic structure inherently experiences complex and bulky configuration for imaging application, which hinders the integration with compact imaging system.

Over the last decade, surface plasmons, i.e., collective oscillation of electrons spatially confined in novel metal nanostructures, have become of much interest owing to their simple structural configuration such as single- or bi-layered nanostructures. In particular, the structural coloration of transmitted light can be achieved by using extraordinary optical transmission (EOT) in the periodic plasmonic nanoholes.[8, 9] According to the EOT, the nanohole arrays (NHA) provide high angular sensitivity for transmitted light, accompanying with high transmission efficiency. This configuration facilitates a simple camera packaging rather than the reflection, and thus the NHA have a great potential for serving as a color filter for high resolution image sensors of submicron pixels.[10-12] However, conventional NHA intrinsically exhibit substantial color-crosstalk due to the multiple transmission peaks under normal incidence of light.[13, 14] Some technical efforts have been made to overcome the color-crosstalk by utilizing index-matching layer deposition,[10, 15] wet-etching,[16] or additional nanolithography of entirely different structures for a specific color,[17] and thus cost-effective fabrication steps are still in need. Furthermore, subsequent emergence of multiple transmission peaks significantly obstructs the structural coloration of high color purity through the NHA as the incident angle increases.[13, 18]

Here we report complementary plasmonic structures (CPS) that exhibit a single resonance peak with high angular sensitivity (**Fig. 1a**). The CPS features a self-aligned configuration of silver nanohole arrays (NHA) and nanodisk arrays (NDA) supported by nanopillars of hydrogen silsesquioxane (HSQ). **Figure 1b** displays the schematic illustration for comparing individual and integrated spectra of NHA and NDA in the CPS, respectively. The NHA show multiple spectral peaks in transmittance spectrum whereas the NDA provide broad transmittance spectrum except a single dip.[19, 20] In contrast, the CPS clearly exhibit a single spectral peak by suppressing a high-order transmission peak of NHA via the structural hybridization of both NHA and NDA. Besides, silver exhibits high Q-factor due to the small imaginary part of material permittivity in the visible and NIR ranges, giving rise to high/sharp transmission peak from the NHA and large extinction from NDA, respectively.[21] As a result, the CPS can serve as a plasmonic filter with high angular sensitivity, which enables continuous tuning of a single transmission peak with high color-purity and Q-factor, depending on an incident angle.

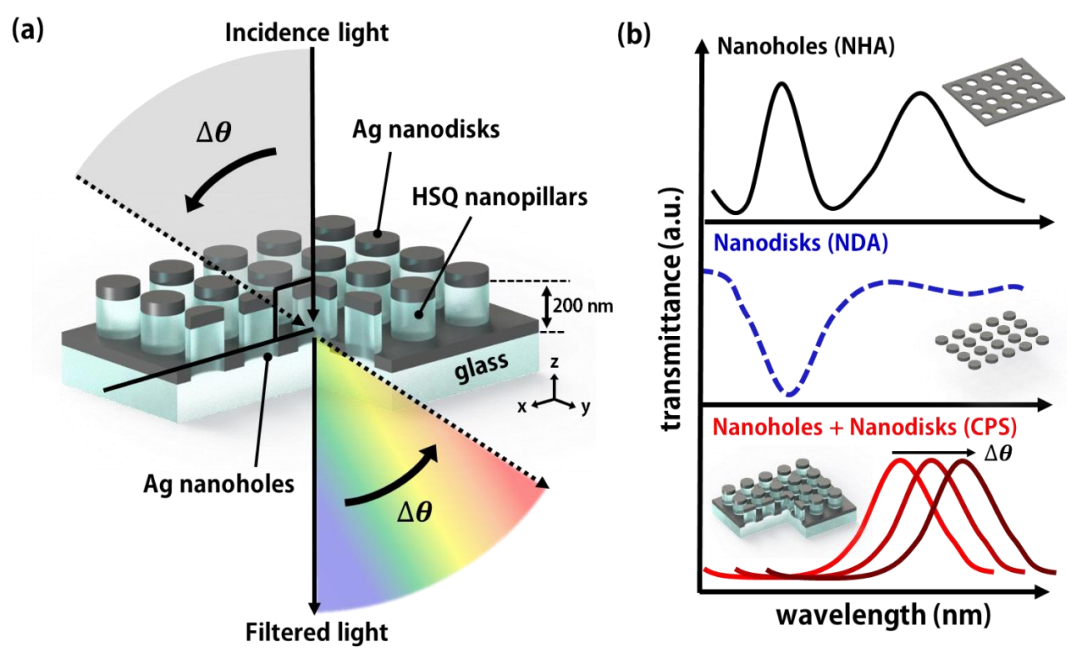

**Figure 1.** (a) Schematic illustration of complementary plasmonic structures (CPS), i.e., self-aligned silver nanohole arrays (NHA) and nanodisk arrays (NDA) that are supported by hydrogen silsesquioxane (HSQ) nanopillars. The CPS selectively transmit a single plasmonic resonance peak (i.e. a single color in the schematics), which becomes red-shifted as the incident angle of light increases. (b) Characteristic spectra of NHA, NDA, and CPS. The second-order transmission peak of NHA is selectively and substantially suppressed by the extinction peak of NDA whereas the first-order transmission of NHA is sustainable in CPS and red-shifted as an incidence angle increases.

The CPS is simply nanofabricated by using electron beam lithography and thermal evaporation as illustrated in **Fig. 2a**. A 200 nm thick HSQ resist (Dow Corning®, XR-1541) is initially spin-coated onto a borosilicate glass substrate and then exposed with electron beams to define nanopillars with smooth vertical sidewalls (Fig. 2b and 2c). The exposed HSQ turns into a glass-like dielectric material, which has a similar refractive index as well as chemical similarities with $SiO_2$.[22, 23] After the development process, the HSQ nanopillars with a slightly negative slope are thermally cured at 150℃ to completely evaporate the solvent.[24, 25] Finally, a thin silver film is thermally evaporated on the nanopillars to construct self-aligned silver nanohole and silver nanodisk arrays. Figure 2c shows that both the NHA and the NDA are apart by the height of HSQ nanopillars and clearly self-aligned on a glass substrate.

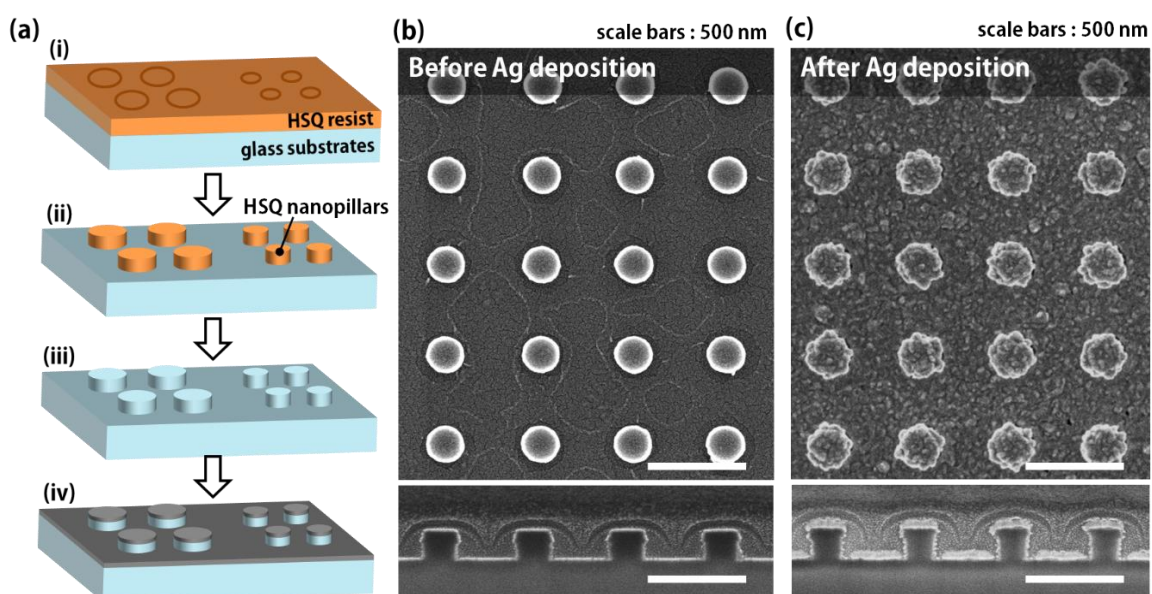

**Figure 2.** (a) Nanofabrication procedures of the CPS. HSQ resist is spin-coated on a glass substrate (i) and exposed with electron beams (ii). HSQ nanopillars are hard-baked and cross-linked to a glass-like dielectric material (iii). Finally, thin silver film is thermally evaporated to form self-aligned silver nanodisk and nanohole arrays (iv). Top and cross-sectional scanning electron microscope (SEM) images of HSQ nanopillars before (b) and after (c) Ag deposition.

Transmission spectra through the NHA, NDA and CPS were numerically calculated by using the three-dimensional finite difference time domain (FDTD) method (**see experimental methods for the details**). **Figure 3a** describes the spectral shifts of NHA (left), NDA (center), and CPS (right), respectively, as the incident angle of light varies from the normal incidence to 30° with respect to the surface normal. The calculated transmittance is displayed on a color scale. The first-order spectral peak, corresponding to the brightest region in the NHA of Fig. 3a, becomes red-shifted as the incident angle increases. In contrast, the second-order spectral peak, observed around a wavelength of 600 nm in normal incidence, is divided into two distinctive peaks as the incident angle increases. This peak separation results from the inherent nature of NHA that all the spectral peaks originating from specific plasmonic resonances (m, n) are split into two distinctive modes (± m, n) as the incident angle increases.[13] In other words, surface plasmons propagating into forward and backward directions have different magnitude of k-vectors under oblique illumination of incident light and thus a spectral peak with resonance (m, n) is split into two different spectral peaks with forward (+m, n) and backward resonances (-m, n). The resonance split of NHA causes multiple spectral peaks at the wavelength of 500 nm to 800 nm when the incident angle is above 15°. In contrast, the NDA exhibits a strong transmission dip (i.e. extinction peak) at around 600 nm in wavelength, which can substantially suppress the unwanted peaks from the NHA. As a result, the complementary configuration of NHA and NDA with a distance of 200 nm selectively attenuate high-order spectral peaks of NHA at the wavelength between 500 nm and 650 nm. The dashed lines (a-a') of **Fig. 3a** are equivalent to the transmission spectra of **Fig. 3b**. For the normal incidence, the CPS clearly exhibit a single peak of plasmon resonance without high-order peaks ranging from 500 nm to 700 nm. The first-order spectral peak of the CPS still remains constant due to the weak interaction with an extinction peak of the NDA, which is drawn as a dashed line. The attenuation of high-order spectral peaks from the NHA is highly associated with the geometric parameters of the CPS such as the period, the diameter, and the thickness. The transmission spectra of the NHA and the NDA were calculated depending on the disk/hole diameter and the Ag thickness, respectively (**see Supplementary information (SI) Fig. S1**). The first-order spectral peak becomes red-shifted from 737 nm to 815 nm in wavelength as the hole diameter of NHA increases from 160 nm to 240 nm under a constant Ag thickness of 50 nm (**Fig. S1 (a)**). The first order spectral peak of NHA is blue-shifted from 821 nm to 775 nm in wavelength as the Ag thickness of NHA increases from 30 nm to 70 nm (**Fig. S1 (b)**). In a similar fashion, the extinction peak of NDA becomes red-shifted from 587 nm to 680 nm in wavelength as the diameter of NDA increases in the same range under a constant Ag thickness of 50 nm (**Fig. S1 (c)**). The extinction peak of the NDA becomes blue-shifted as the Ag thickness increases from 30 nm to 70 nm (**Fig. S1 (d)**). The calculated results show the first-order transmission peak of NHA and the extinction peak of NDA are substantially shifted whereas the second-order peak of NHA is slightly shifted despite the geometrical change. As a result, it clearly shows that the second-order spectral peak is effectively attenuated by controlling the extinction peak, depending on the disk diameter and the Ag thickness. For quantitative analysis, the degree of attenuation (DOA), i.e., a ratio of the attenuated intensity to the original intensity for the second-order peak of NHA, were numerically calculated with two geometric parameters of disk/hole diameter and silver thickness (**Fig. 3c**). High DOA indicates a significant attenuation of the second-order peaks. The specific dimensions for high DOA at the normal incidence are plotted as the bright region as shown in **Fig. 3d**. The dark region is also referred to low DOA at normal incidence, which indicates the second-order peak is barely attenuated through the CPS. The inset figures of **Fig. 3d** show the calculated cross-sectional electric field distributions of low DOA (DOA = 0.63) and high DOA (DOA = 0.96), respectively for the corresponding wavelength of second-order spectral peak from NHA ($\lambda_{NHA2}$). The electric field distribution explains that the DOA is strongly associated with the volumetric electromagnetic (EM) field confinement inside the HSQ nanopillars. For instance, the EM field is strongly confined within the HSQ nanopillars at high DOA (DOA = 0.96) whereas the weak confinement of EM field is observed at low DOA (DOA = 0.63). The calculated result implies that the strong EM field confinement within HSQ nanopillars suppresses the transmission of light at $\lambda_{NHA2}$. The second-order transmission peak of NHA is remarkably attenuated when the wavelength separation between $\lambda_{NHA2}$ and $\lambda_{NDA}$ (i.e. a wavelength of the extinction peak from NDA) is approximately 30 – 40 nm, in case that the $\lambda_{NDA}$ is longer than $\lambda_{NHA2}$ (**Fig. S2 (a)**). The value of DOA decreases unless maintaining the wavelength separation as shown in **Fig. S2 (b)-(c)**. The bright region in DOA map would gradually shift into bottom-right direction as the incident angle increases from normal to 30°. Consequently, the value of DOA depending on the Ag thickness and disk/hole diameter directly provides the functional operating window of CPS.

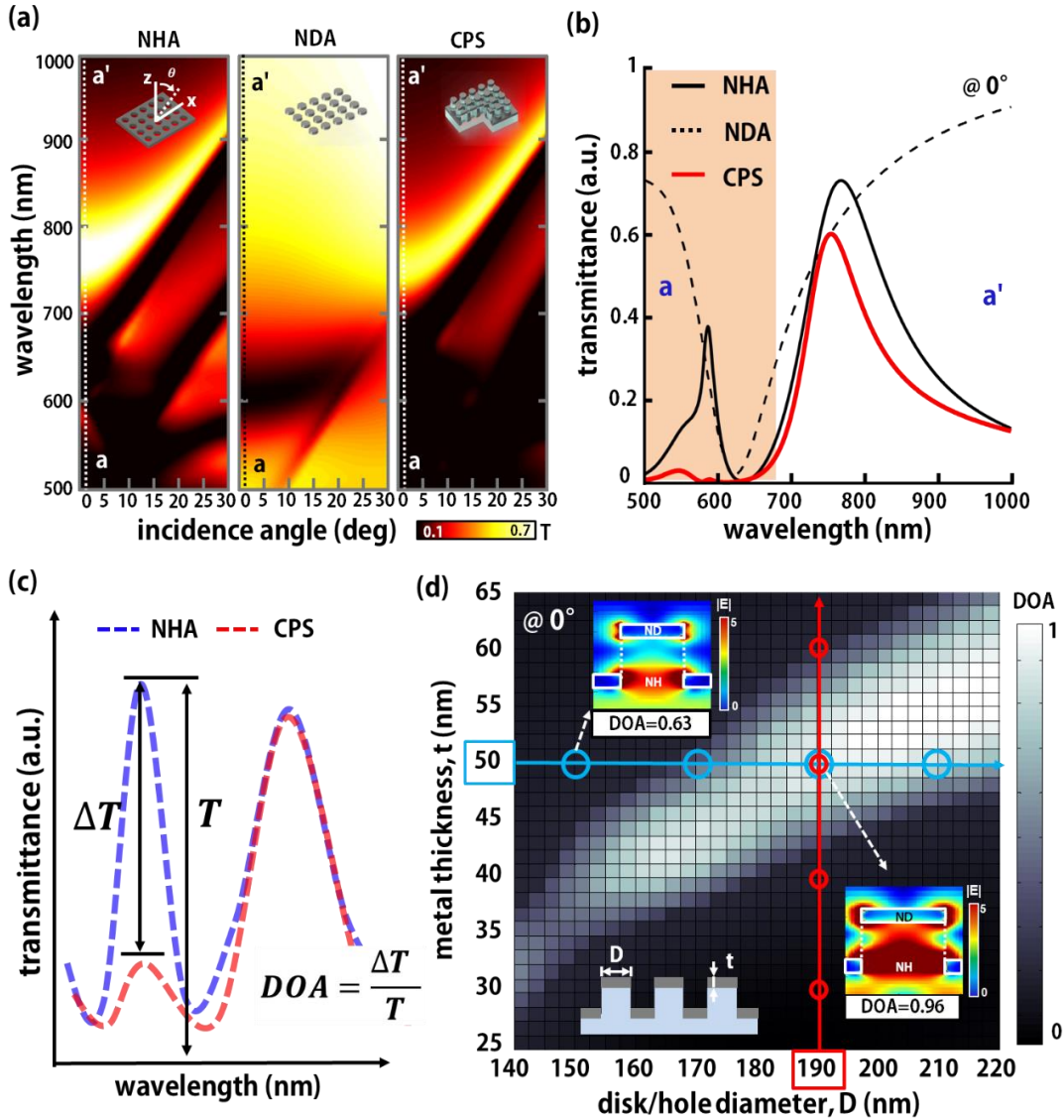

**Figure 3.** (a) Numerical calculations of spectral shifts for NHA, NDA, and CPS depending on the incident angle of light. The first-order spectral peaks from NHA and CPS are similarly shifted as the incident angle is increased from 0° to 30° whereas the high-order peaks from NHA are substantially attenuated in the range of 500 nm to 650 nm. (b) The transmission spectra of NHA, NDA, and CPS are extracted at the incidence of 0°. The spectra of CPS clearly show the attenuated high-order peaks of NHA due to the extinction of NDA, which serve as a spectral suppressor. (c) A simplified illustration of the degree of attenuation (DOA), which is defined as a ratio of the attenuated intensity to the original intensity of NHA. (d) DOA mapping as a function of the disk/hole diameter (nm) and Ag thickness (nm) at the normal incidence. The DOA is directly controlled by the hole/disk diameters and Ag thickness. The inset figures indicate the calculated cross-sectional electric field distributions at low DOA (top left) and high DOA (bottom right), which imply that the attenuation of transmitted light at the $\lambda_{NHA2}$ results from the EM field confinement inside HSQ nanopillars.

The second-order spectral peak of CPS sample was experimentally measured by using an inverted microscope (Carl Zeiss, Axiovert 200M) equipped with a spectrometer (Princeton Instruments, MicroSpec 2300i). **Figure 4a** shows the calculated (dashed line) and measured (solid line) transmittance at a corresponding wavelength for the second-order peak of CPS ($\lambda_{CPS2}$) depending on the disk/hole diameters and metal thickness. The red lines correspond to transmitted intensities at $\lambda_{CPS2}$ when the metal thickness is varied from 30 nm to 60 nm under the constant disk/hole diameter of 190 nm. The measured results (red solid line) demonstrate the lowest intensity for 50 nm in Ag thickness and the highest intensity for 30 nm, which have a similar tendency with the calculated results shown as red dashed line. In a similar fashion, the blue lines show the intensities at $\lambda_{CPS2}$ depending on the disk/hole diameters under the constant Ag thickness of 50 nm. The measured transmittance (blue solid line) decreases as the disk/hole diameter increases from 150 nm to 190 nm and subsequently increases as increments up to 210 nm. The measured results agree well with the calculated results shown as the blue dashed line. The experimental results also show that the substantial attenuation of the second-order spectral peaks is clearly observed when the disk/hole diameter and the metal thickness are 190 nm and 50 nm, respectively (**see Supplementary information (SI) Fig. S3 for the detailed spectra**). The transmission spectra of CPS with the disk/hole diameter of 190 nm and the Ag thickness of 50 nm were measured depending on the incident angle ranging from normal to 30° (**Fig. 4b**). The experimental results clearly demonstrate the wavelength of first-order spectral peak ($\lambda_{CPS1}$) becomes red-shifted from 731 nm to 843 nm, where a distinctive single peak is shown along all the incident angles. The full width at half maximum (FWHM) of transmission peaks decreases from 120 nm to 80 nm as the incident angle increases from normal to 30°, which provides the CPS with high selectivity for specific wavelengths, compared to conventional plasmonic filters.[10, 12] **Figure 4c** shows each $\lambda_{CPS1}$ corresponds to each incident angle increasing by 5° from 0° to 30°, which clearly demonstrates that the first-order spectral peak of CPS has a monotonic relationship with the incident angle. The relationship, i.e. spectral tuning capability of CPS, was quantified by angular sensitivity (AS) of the first-order spectral peak, which is defined as the spectral shift of $\lambda_{CPS1}$ over a total incident angle change (0°- 30°). According to the Fig. 4c, high AS of 4.8 nm/deg was achieved at 450 nm period in CPS, referring to wide tuning range, while the small AS of 2.0 nm/deg and narrow tuning range was achieved at 350 nm period in CPS. The measured results support that the AS increases with the period of CPS, which is well matched with the calculated results (**Fig. S4**). In addition, spectral region of CPS (i.e. the region where all the $\lambda_{CPS1}$ are located) responds to the change of periods. For instance, the spectral region is heavily biased toward the NIR range in case of the CPS with 450 nm period. The CPS, tailoring the tuning capability by array period, facilitates the reliable and predictable spectral tuning of a single spectral peak depending on the angle of incident light.

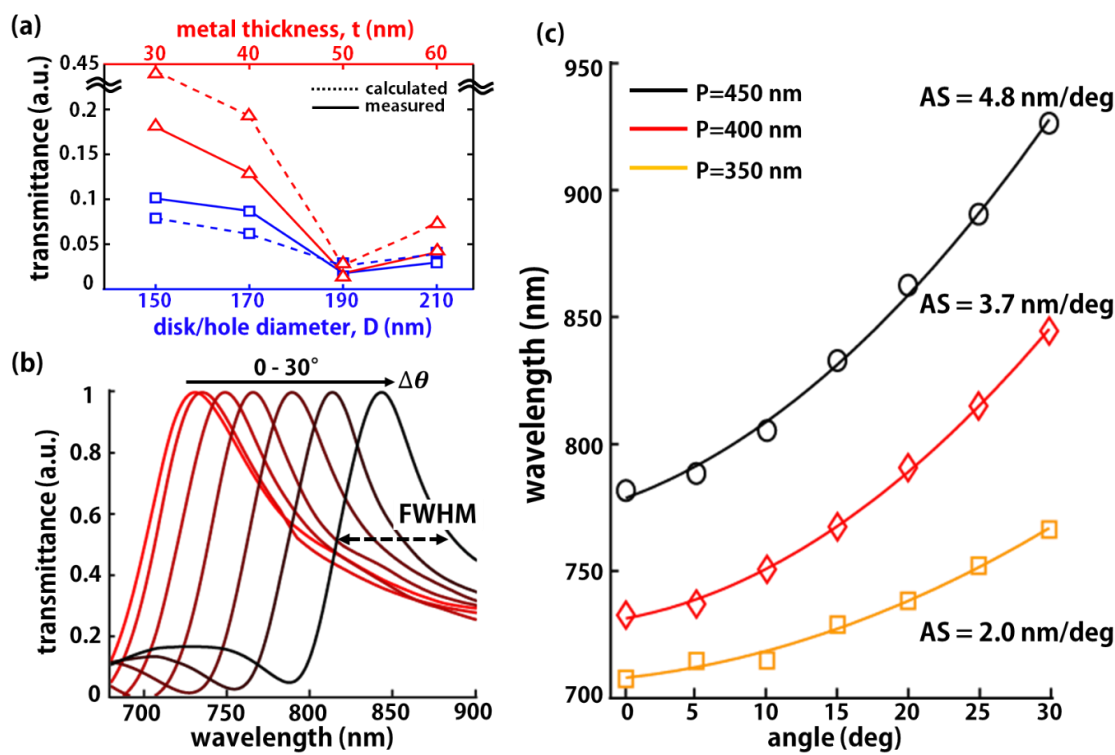

Figure 4. Spectral measurements for the second-order spectral peak transmittance and angle-sensitive tunability of CPS. (a) Peak transmittance at $\lambda_{CPS2}$ depending on the hole/disk diameter and the Ag thickness. The peak transmittance is highly attenuated for the CPS of D=190 nm and t=50 nm. (b) Measured transmittance spectra depending on the incident angle. The first-order spectral peak is red-shifted as the incident angle increases. (c) Measured $\lambda_{CPS1}$ depending on the incident angle and the period of CPS. The CPS with large array period can provide a high AS and longer wavelength range for spectral tuning.

In summary, the structural coloration of transmitted light has been successfully demonstrated by using the structural hybridization of self-aligned and metallic nanodisk and nanohole arrays, which effectively attenuate the unwanted second-order spectral peak of NHA. A single transmission peak of 80 nm in FWHM is tuned from 731 nm to 843 nm in wavelength as the incident angle of light increases from 0° to 30°. Therefore, The CPS can be a promising candidate for a tunable filter with high color-purity by virtue of high DOA and narrow FWHM. This novel spectral filter with high tunability can play an essential role for compact multispectral or hyperspectral cameras or microspectrometer.

ASSOCIATED CONTENT

**Supporting Information**.

Experimental methods and figures (S1-S4) showing supplementary calculated/measured data (PDF)

AUTHOR INFORMATION

**Corresponding Author**

*E-mail: kjeong@kaist.ac.kr

**Author Contributions**

M.-S. A. and K.-H. J. conceived and conducted the experiments. M.-S. A. and T. C. analyzed the results. The manuscript was written and reviewed through contributions of all authors. All authors have given approval to the final version of the manuscript.

**Notes**

The authors declare no competing financial interest.

ACKNOWLEDGMENT

This work was supported by Samsung Research Funding Center of Samsung Electronics under Project Number SRFC-IT1402-02.